\documentstyle[prl,aps,multicol]{revtex}
\begin{document}
\draft
\title{Propagation of Coulomb-correlated
electron-hole pairs in semiconductors with correlated and 
anticorrelated disorder}
\author{P.~Thomas$^1$, I.~Varga$^{1*}$,
T.~Lemm$^{1}$, J.E.~Golub$^1$, K.~Maschke$^2$, T.~Meier$^1$, S.W.~Koch$^1$}
\address{$^1$Fachbereich Physik und wissenschaftliches Zentrum f\"ur
Materialwissenschaften,
Philipps-Universit\"at Marburg,
Renthof 5, D-35032 Marburg, Germany}
\address{$^2$Institut de Physique Appliqu\'ee,
\'Ecole Polytechnique F\'ed\'erale de Lausanne, CH-1015 Lausanne,
Switzerland}

\maketitle

\begin{abstract}
Local ultrafast optical excitation of electron-hole pairs in disordered
semiconductors provides the possibility to observe experimentally
interaction-assisted
propagation of correlated quantum particles in a disordered
environment. In addition to the interaction driven delocalization
known for the conventional single-band
TIP-(two-interacting-particles)-problem 
the semiconductor model has a richer variety of 
physical parameters that give rise to new features in the temporal dynamics.
These include different masses, correlated vs.
anticorrelated disorder for the two particles, and dependence on
spectral position of excitation pulse. 
\noindent
\end{abstract}
\begin{multicols}{2}
\narrowtext
\noindent
{\bf Introduction.}
It is by now well established that the center-of-mass (COM) motion of
two interacting particles (TIP) in a 
single tight-binding band of a one-dimensional
Hamiltonian possessing diagonal disorder
is spatially more extended than both their relative motion
and the motion of the particles without interaction \cite{shepel}.
Although this model is a fundamental paradigm for the more general
question of the interplay of disorder and interactions in many-body
systems, it suffers from not being susceptible to experimental
verification. Furthermore, it is not clear whether this finding,
relevant for two particles, can be
generalized to real transport situations in dense Fermi systems.

On the other hand, low-intensity optical excitation in disordered
semiconductor systems produces strongly Coulomb-correlated electron-hole 
pairs with large mutual separation such that their motion can be
considered to be that of an essentially isolated pair. Optical
pump-probe experiments are in principle suited to record the 
coherent dynamics
of the pair after pulsed excitation on a ps-time scale before scattering
with acoustic phonons becomes relevant. 

Following this idea
the temporal traces of the participation number of the electron and the hole
have been studied in a previous paper \cite{dirk} by solving the
equation of motion for the two-particle amplitude $p_{ij}(t)=\langle
\hat d_i \hat c_j \rangle$, which is
the interband coherence related to the optical polarization. Here
$\hat d_i$ ($\hat c_j$) are hole (electron) operators at site $i$ ($j$).
In the low intensity (with respect to the exciting light pulse)
limit the sum rules  $n_{ij}^e=\sum_l p_{lj}p_{li}^*$
and $n_{ij}^h=\sum_l p_{jl}p_{il}^*$ yield the electron (hole)
intraband quantities. Their diagonal elements $n_{ii}^{e,h}$ are the
time dependent densities of electron and hole. Defining the participation
number $\Lambda(t)=(\sum_i n_{ii}^2)^{-1}$ the evolution of the excited
electron and hole wave packet can be plotted as a function of time. 
Alternatively, the COM-coordinate $R$ and the relative
coordinate $\rho$ of the electron-hole wave packet
can be calculated directly 
from the pair amplitudes.

It has been demonstrated that the interaction induces an
enormous change of the dynamics if compared to the non-interacting
case. Instead of an exponential rise of $\Lambda(t)$ towards a
saturation value on the time scale of the excitation pulse, in the interacting
case a slow, diffusion-like rise of $\Lambda$ is seen that does
not seem to saturate in the limited time regime accessible to
the numerical calculation (and relevant for the real physical
situation). For brevity this feature will be called
``enhancement'' in the following in accordance with
the notion of other work in the field,
although it is not implied that from the present 
dynamical calculation in a finite
time domain
anything like the enhancement of a localization length can be
deduced. The dependence
of the enhancement on the sign of the interaction has been studied and
it was found that  the dependence on the sign vanishes 
only for excitation in the center of the 
continuum. 

In this contribution the effect of correlated versus anticorrelated
disorder and of the interaction strength is studied.

{\bf Model.}
A two-band Hamiltonian $H_0$ is considered
\begin{eqnarray}
H_0&=&\sum_{i=1}^N(\epsilon_i^e \hat c_i^{+} \hat c_i +
\epsilon_i^h \hat d_i^{+} \hat d_i)
-J^e\sum_{i=1}^N(\hat c_i^+\hat c_{i+1} + \hat c_{i+1}^+c_i)\nonumber \\
&-&J^h\sum_{i=1}^N(\hat d_i^+\hat d_{i+1} + \hat d_{i+1}^+d_i)
\end{eqnarray}
with nearest neighbour coupling $J^e$ ($J^h$) for electrons (holes) 
and diagonal disorder given by a box-shaped distribution function
of the single-site energies $\epsilon_i^e$ ($\epsilon_i^h$) of total width
$W$. The $N$ sites form a linear chain with spacing $a$ and periodic
boundary conditions. 

The interaction is given by a regularized Coulomb potential 
\cite{laci} in
monopole-monopole form
\begin{equation}
H_C=\frac{1}{2}\sum_{i,j=1}^N (\hat n_i^e-\hat n_i^h)V_{ij}(\hat n_j^e-n_j^h)
\end{equation}
with
\begin{equation}
V_{ij}=-\frac{U}{4 \pi \varepsilon \varepsilon_0}\frac{e^2}{r_{ij}+\alpha}
\end{equation}
where $\alpha=5a$ and $U$ is a 
dimensionless parameter quantifying both strength and sign
of the interaction.

The optical excitation is given by a local dipolar coupling to the light field
\begin{equation}
H_I=-\sum_{i=1}^N(\mu_i E^*(t) \hat d_i \hat c_i + h.c.).
\end{equation}
For local excitation at site $0$ the optical dipole matrix elements are
taken to be $\mu_i=\mu_0\delta_{i,0}$.
The light field is given by
\begin{equation}
E(t)=(\pi)^{-1/2}\sigma^{-1}\exp{(-(t/\sigma)^2)}\exp{(-i\omega_0t)}
\end{equation}
with central frequency $\omega_0$ and temporal width $\sigma$. 

{\bf Equation of Motion.}
The equation of motion for the pair amplitude $p_{ij}$ is derived
using the total Hamiltonian $H=H_0+H_C+H_I$. In the low intensity limit
linear response theory is valid and the equation of motion reads
(with $\hbar=1$)
\begin{eqnarray}
\partial_t p_{ij}&=&-i(\epsilon_i^e+\epsilon_j^h-V_{ij})p_{ij}\nonumber \\
&+&i\sum_{l=1}^N(J^ep_{il}+J^h
p_{lj}) + i\mu_j E(t)\delta_{ij}.
\end{eqnarray}
This equation of motion is solved numerically for $M$ realizations 
(typically $M=20...40$) of the
disorder drawn from the distribution of site energies. The observables
are then configurationally averaged over these realizations.

{\bf Choice of Parameters.}
As the present model describes a situation that is in principle 
accessible to experiments, the interaction strength is
chosen to yield exciton binding energies in the range of values
typical for low-dimensional semiconductor heterostructures. The
intraband couplings $J^{e,h}$ and the strength of disorder $W^{e,h}$
are free parameters. However, the tight-binding model 
is thought to model the conduction and valence band extremities of
a disordered semiconductor heterostructure. 
In this communication we show data for $J^e=J^h=20meV$
exclusively, i.e. equal electron and hole masses are assumed.
The lattice constant, equal to a disorder length scale, 
is taken to be $a=20 \AA$.
The external
optical pulse resembles typical laser pulses used in ultra-fast
optical experiments on coherent phenomena. 
Its central frequency $\omega_0$ 
is chosen to be situated below the center of the
absorption band calculated 
without interaction, in order to model the dynamics of
electron-hole pairs in the continuum in a range of energies that
excludes LO-phonon emission. 
Specifically, the gap of the noninteracting case is the
origin of the energy axis, 
the band center is at $\Delta=80$ meV,
and $\omega_0= 40$ meV.
For this excitation condition the sign
of the interaction is not irrelevant \cite{dirk}. Here we use attractive
interaction throughout. A temporal pulse width
$\sigma=$100 fs is used corresponding to a spectral width 
of 22 meV.
The number of sites $N$ is taken
large enough (typically $N=240$) such that for not too small
disorder the locally excited wave packet does not reach the 
boundary of the sample in typically some ten picoseconds. After this
time acoustic phonon scattering will lead to effective dephasing and the
coherent phenomena studied here will be destroyed. For principal
studies, however, the calculation is sometimes performed for much longer times,
although the results are no longer relevant to experiments.

{\bf Observables.}
The extension of the electron and hole wave packets as a function of time 
can be characterized by the above mentioned participation number
$\Lambda^{e,h}$ which are single-particle quantities. 
Using the pair amplitude $p_{ij}$, the two-particle quantities
COM-coordinate
$R=(\sum_{i,j}|p_{ij}|^2(i+j)^2/2)^{1/2}$ and relative coordinate
$\rho=(\sum_{i,j}|p_{ij}|^2(i-j)^2/2)^{1/2}$ can be defined. Their
ratio $R/\rho$ is a measure of the interaction induced enhancement.

{\bf Correlated and Anticorrelated Disorder.}
While previous results \cite{dirk} have been obtained using uncorrelated
disorder, here the influence of correlation is studied. Disorder 
for electrons and holes is called 
correlated if the site energies of electrons 
equal that of the mirror image of the hole energies
with respect to the gap center. This kind of
disorder is expected in heterostructures due to
interface roughness leading to a spatially
fluctuating
confinement of single-particle wave functions.
On the other hand, if the 
energy separation within an isolated site remains constant, i.e.
equal to $\Delta$, the disorder
is called anticorrelated. 
A spatially fluctuating electrostatic field would induce this kind of
disorder.
The present model resembles that of the
conventional single-band-TIP if electrons and holes have equal masses 
and
if correlated disorder (with $W^e=W^h$) 
is applied. Electron and hole
wavefunctions are then pairwise equal, both particles live in the same
environment. For $U=0$ all optical matrix elements connecting these
pairs of states are equal, all others are zero. Therefore,
the optical spectrum for $U=0$
resembles the density of single-particle states, however, with
bandwidth being the sum of that of the two bands. 
As in the last years the conventional TIP-model has been widely studied
we here concentrate mainly on the anticorrelated situation.

Figure \ref{fig1} shows COM-traces for correlated and anticorrelated
disorder for $U=0$ and $U=3$. Remarkably, the enhancement for anticorrelated
disorder is much more pronounced compared to the correlated case.
These traces have been calculated for an optical excitation energy
$\omega_0 = 40$ meV. Looking at the optical spectra
for the anticorrelated case, Fig. \ref{fig2},
one realizes a peak that for $U=0$ lies in the center of the band at 
$\Delta=80$ meV and shifts towards lower energy for increasing (attractive)
$U$. For the excitation condition leading to the large COM-enhancement
at $U=3$ pairs of states in the vicinity of
this peak are excited. The origin 
of the peak can be traced back to optical transitions
connecting states in the tails of the single-particle bands.
Even for moderate disorder these
pairs of states with the nearly identical transition energy 
are strongly localized 
at the same position, and therefore
their optical matrix elements are large.
 Exact eigenvectors for a short sample ($N=10$) have
been calculated for $U=0$ and $U=3$, confirming the strong 
overlap
within the contributing pairs. Obviously,
these strongly localized tail states are not responsible
for the large enhancement. 

The fact that the enhancement for anticorrelated disorder is larger than
that for correlated disorder is not fully understood yet. A prominent
difference exists in the equations of motion for $p_{ij}$ for the two
cases. For the anticorrelated case the disorder is absent in the
equation for the diagonal elements $p_{ii}$, while it is present
in the respective equation for the correlated case. 

The behavior of excitons in disordered systems is often 
\cite{sergei}
discussed in terms
of the relative coordinate 
$r=i-j$ and the COM-coordinate $x=(i+j)/2$ instead of the indices
$i$ and $j$ (for equal masses,
not to be confused with $R$ and $\rho$). 
After the transformation
$(i,j) \rightarrow 
(r,x)$ the relative coordinate $r$ can be integrated out 
in the equation of motion if, e.g., 
the disorder $W$ is smaller than the exciton binding energy and if
discrete excitonic resonances are considered. 
In our case the equation of motion 
in terms of $x$ would not contain any disorder in
the anticorrelated case, in contrast to the correlated case. Although
for the parameters used in Fig. \ref{fig1} this approach is not strictly valid,
it suggests a possible solution for the problem at hand.

{\bf Finite-Time Scaling.}
In order to quantify the enhancement finite-time scaling
is applied in the following sense. Although with a dynamical
calculation for finite times
it is impossible to decide whether the two-particle packet is
localized or not a long-time
saturation value of the COM-coordinate $R_{\infty}$ 
is assumed
for practical purposes. While the temporal rise of the
interaction-free traces is exponential with time scale given by that
of the pulse duration, the short time behavior in the interacting
case looks like a diffusive process. So the following interpolation
relation is used:
\begin{equation}
R(t)=((Dt)^{-1/2}+R_{\infty}^{-1})^{-1}.
\end{equation}
Figure \ref{fig3} shows the very good quality of typical fits for
times larger than the pulse width. It is found that while the
diffusivity $D$ is weakly dependent on disorder and interaction, 
$R_{\infty}$ shows a much stronger dependence. In the following $R_{\infty}$
is taken to characterize the enhancement. 

{\bf Contour Plots.}
Figure \ref{fig4} shows contour plots of the
two-particle wave function, 
$|p_{ij}|^2$, for anticorrelated and correlated
disorder at time $t=165$ ps. All other parameters are equal to that 
used in Fig. \ref{fig1}. 
Again, the anticorrelated situation shows much larger
enhancement.

{\bf Dependence of Enhancement on Interaction Strength.}
Recently the dependence of the enhancement on interaction $U$
has been discussed and a duality relation 
describing the behavior for large and small $U$
has been proposed \cite{wer}. The
present calculations yield similar results, see Fig. \ref{fig5}
for anticorrelated disorder. The
ratio $R_{\infty}/\rho_{\infty}$ 
is close to unity for small $U$, while it has
a maximum around $U=3...4$ that increases with increasing disorder. Of
course, both $R_{\infty}$ 
and $\rho_{\infty}$ decrease with increasing disorder, however,
$\rho_{\infty}$ faster than $R_{\infty}$. 
Remarkably, the position of the maximum does not depend on
disorder. 
This behavior of
$R_{\infty}/\rho_{\infty}$ is apparently related to the fact that for the
chosen excitation energy $\omega_0$ the center of the 
optical spectrum, indicated by the above mentioned peak,
shifts (c.f. Fig. 2) through the spectral position of the
laser pulse with changing $U$. Around 
$U=3...4$ the peak coincides with the excitation
energy $\omega_0$, while at larger $U$ the excitation takes place again 
in a region where states outside the center of the single-particle
bands are optically coupled. It is known that the enhancement is more
pronounced for less localized single-particle states \cite{shepel,dirk}.
These states are situated in the center of the single-particle
bands.  The Coulomb enhancement
is therefore largest if the
electron-hole packet is generated 
in the center of the optical spectrum since the interaction
couples these weakly localized states most effectively.

{\bf Conclusions.}
For a situation that is in principle accessible to ultra-fast
coherent optical experiments, the delocalizing action of
the Coulomb interaction on optically generated electron-hole pairs has been
studied. The temporal traces of the center-of-mass coordinate
$R$ and the relative coordinate $\rho$ have been
fitted by an interpolation scheme describing  diffusive 
and localized
behavior at small and large times, respectively. 
It should be noted, however,
that the calculation presented here is unable to yield definite answers about
localization or otherwise.
Correlated and anticorrelated disorder is studied and
it is found that the enhancement of $R_{\infty}/\rho_{\infty}$ 
is much more pronounced for anticorrelated
disorder.

\noindent
{\bf Acknowledgements}
This work is supported by DFG, SFB 383, the Leibniz Prize,
OTKA (T029813, T024136, F024135), 
SNSF (2000-52183.97), and the A. v. Humboldt Foundation.

\begin{figure}[ht]  
\begin{center}
\caption{Center-of-mass $(R)$ traces for $U=0$ and $U=3$
for anticorrelated and correlated
disorder. Disorder $W^e=W^h=W=60$ meV.}
\label{fig1}
\end{center}
\end{figure}

\begin{figure}[ht]  
\begin{center}
\caption{Absorption spectra for $U=0$ and  $U=3$ for correlated (dotted
line)
and anticorrelated (solid line)
disorder. Other parameters as in Fig. 1. }
\label{fig2}
\end{center}
\end{figure}

\begin{figure}[ht]  
\begin{center}
\caption{Fit of the center-of-mass trace by Eqn. (7)
for different interaction strengths $U$. Other
parameters are the same as in Fig. 1.}
\label{fig3}
\end{center}
\end{figure}

\begin{figure}[ht]  
\begin{center}
\caption{Contour plots of $|p_{ij}|^2$ at 165 ps
for anticorrelated (left) and correlated (right)
disorder. $U=3$. Other
parameters are the same as in Fig. 1.}
\label{fig4}
\end{center}
\end{figure}

\begin{figure}[ht]  
\begin{center}
\caption{Dependence of the enhancement $R_{\infty}/\rho_{\infty}$
on interaction strength 
$U$ for three different disorder strengths 
$W^e=W^h=W$. Inset shows $R_{\infty}$ (full symbols)
and $\rho_{\infty}$ (open symbols) vs. $U$ separately.}
\label{fig5}
\end{center}
\end{figure}
\end{multicols}
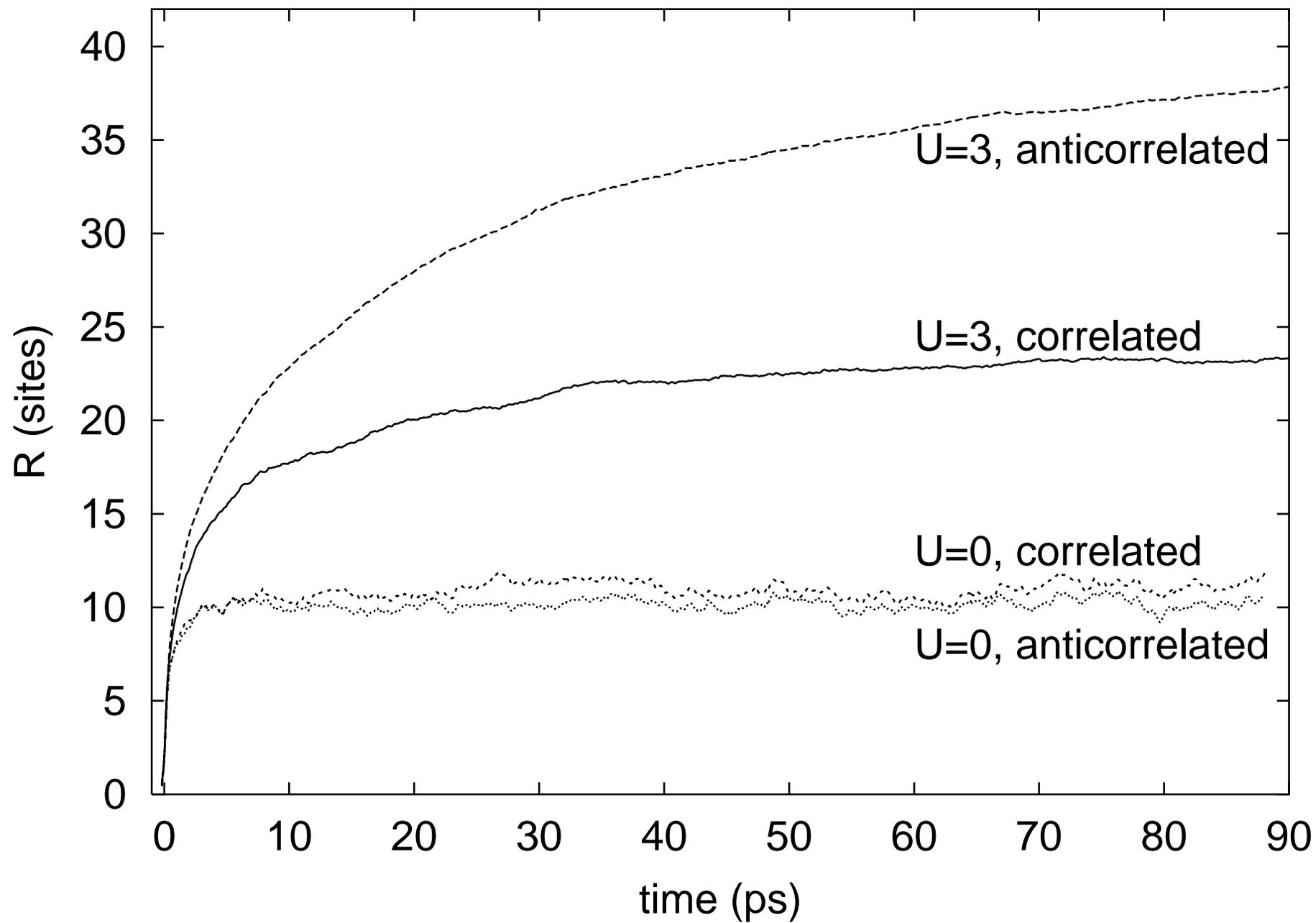
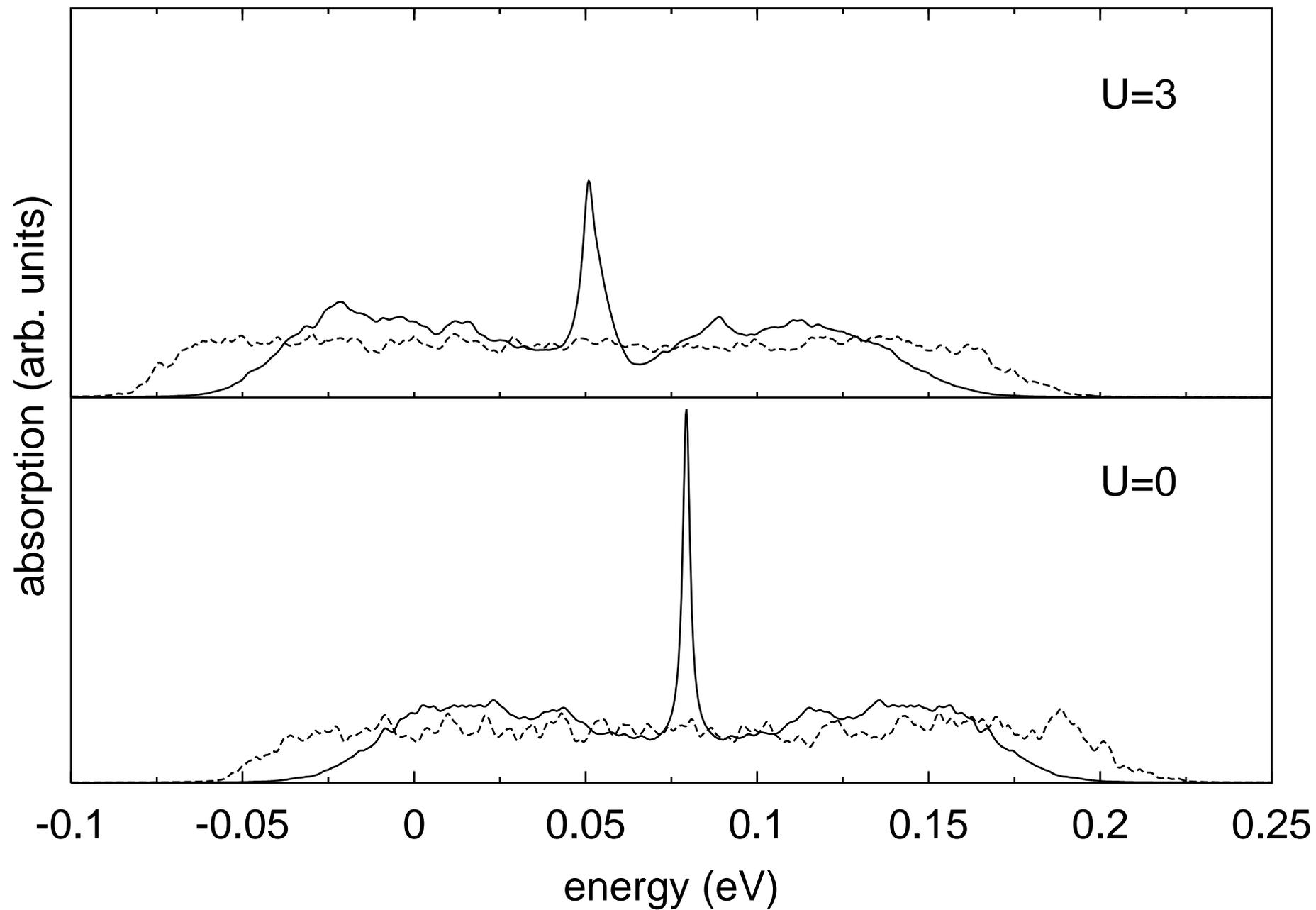
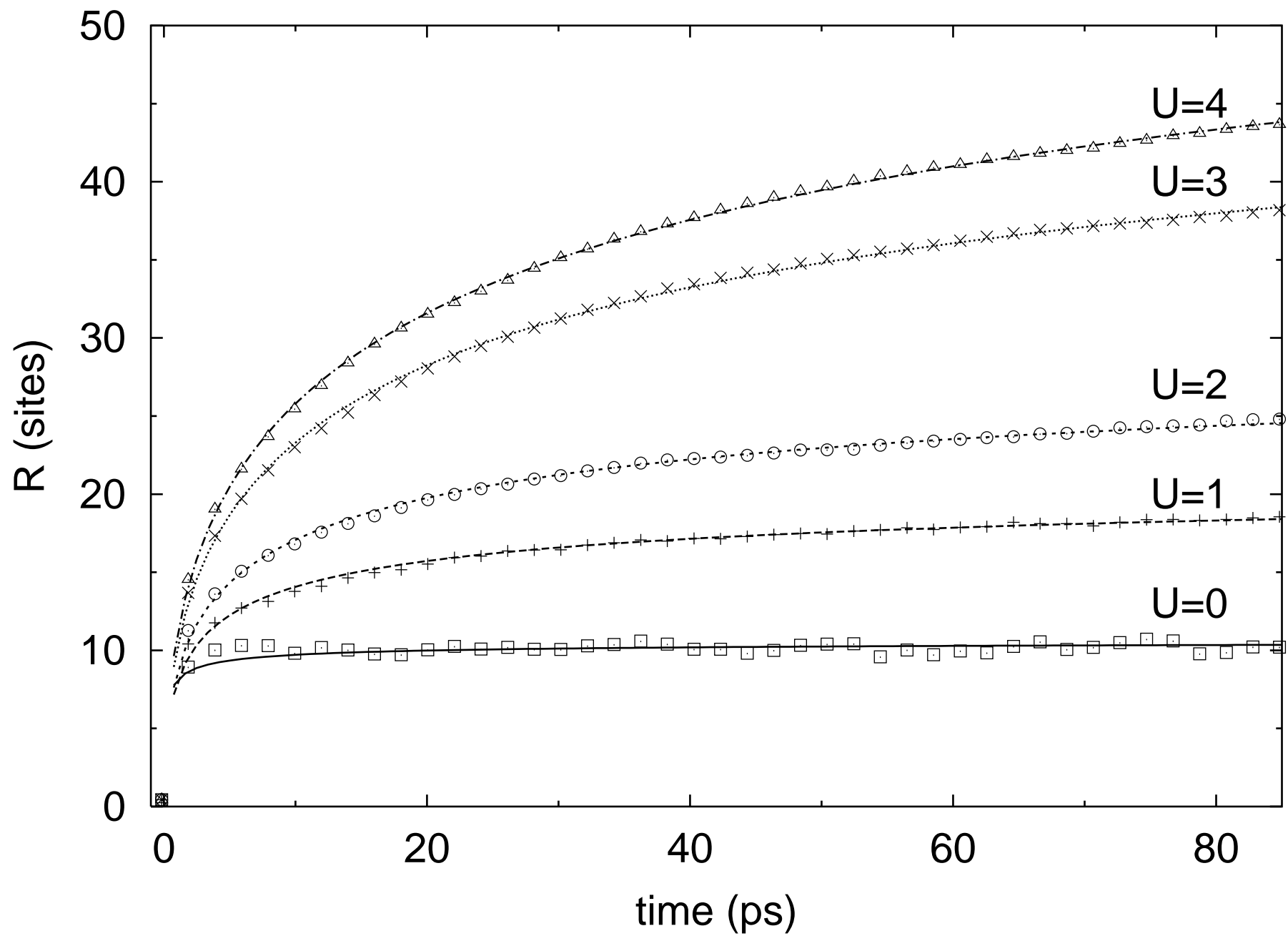
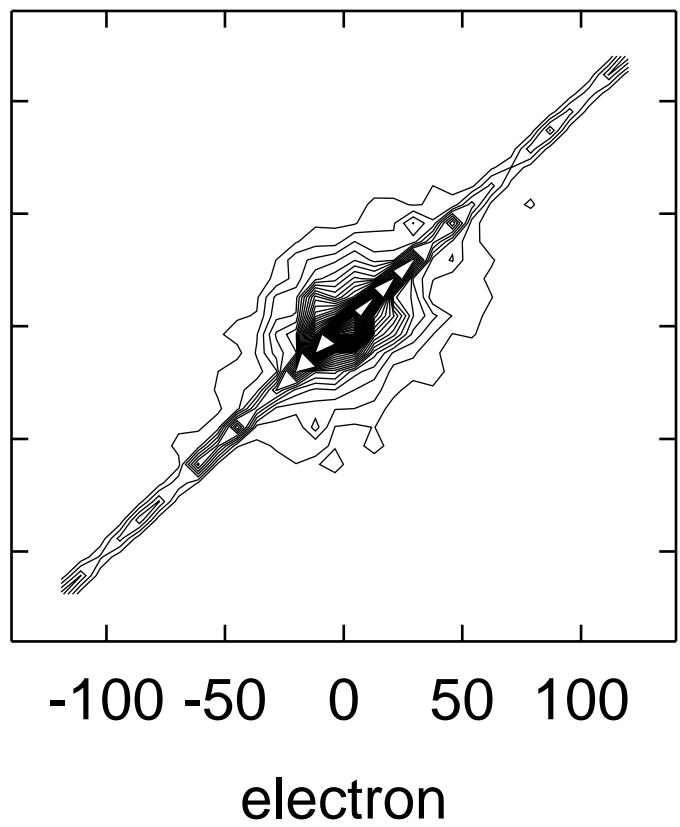
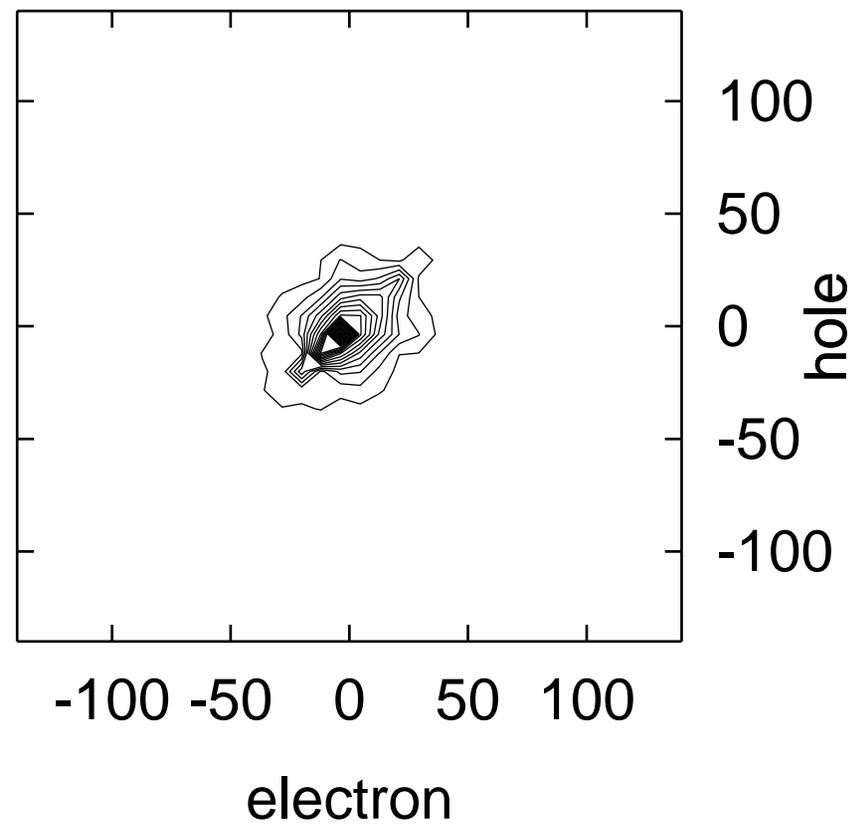
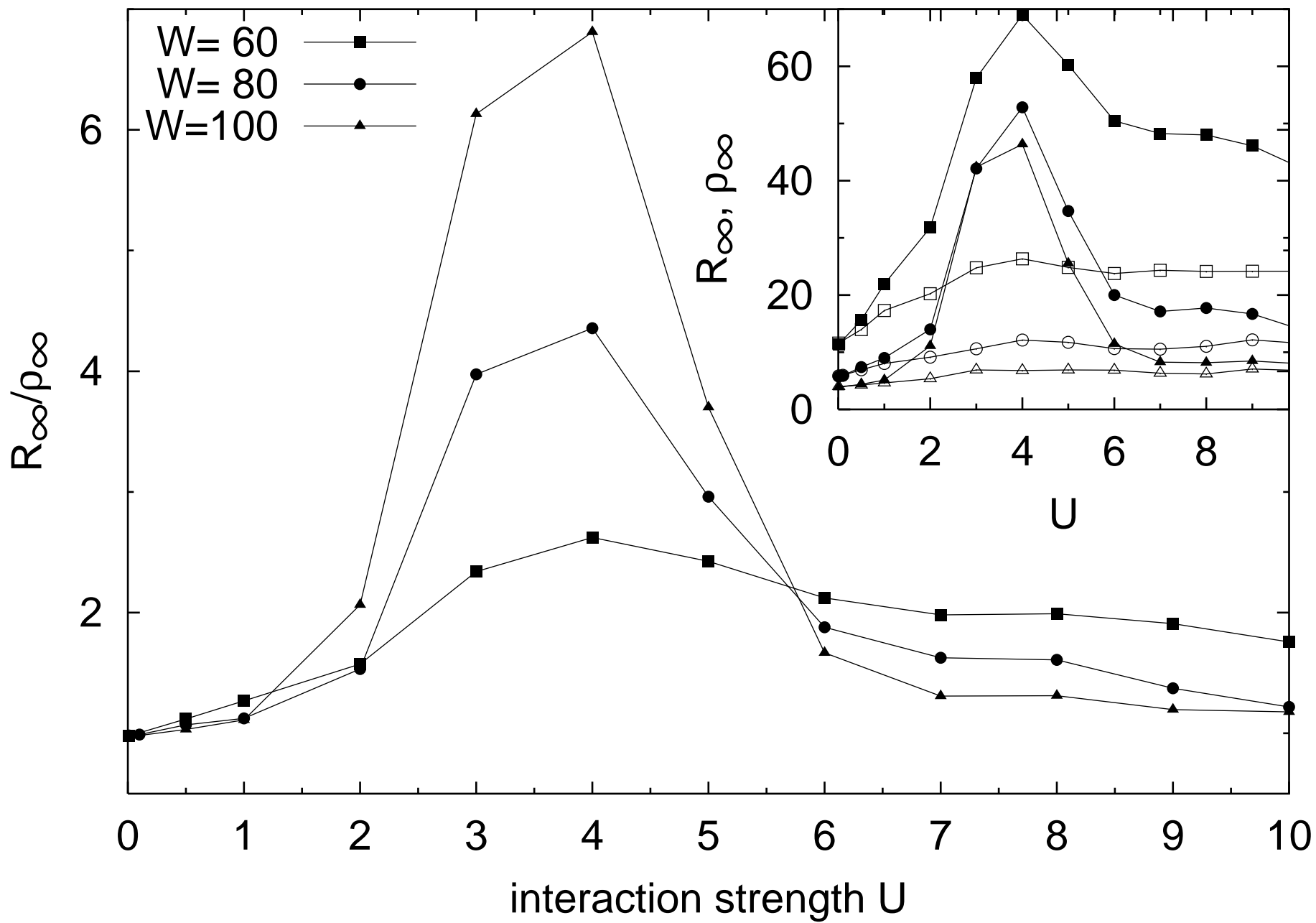
\end{document}